\documentclass[acmlarge]{acmart}
\usepackage{array}
\AtBeginDocument{%
  \providecommand\BibTeX{{%
    \normalfont B\kern-0.5em{\scshape i\kern-0.25em b}\kern-0.8em\TeX}}}

\setcopyright{acmcopyright}
\copyrightyear{2022}
\acmYear{2022}
\acmDOI{XXXXXXX.XXXXXXX}

\acmJournal{JOCCH}
\acmVolume{0}
\acmNumber{0}
\acmArticle{0}
\acmMonth{0}

\begin{document}

\title{CLEF. A Linked Open Data native system for Crowdsourcing}

\author{Marilena Daquino}
\email{marilena.daquino2@unibo.it}
\orcid{XXXXXXXX}
\affiliation{%
  \institution{University of Bologna}
  \streetaddress{via Zamboni 32}
  \city{Bologna}
  \state{Italy}
  \country{Italy}
  \postcode{40126}
}

\author{Mari Wigham}
\affiliation{%
  \institution{The Netherlands Institute for Sound and Vision}
  \city{Amsterdam}
  \country{Netherlands}
}
\email{mwigham@beeldengeluid.nl}

\author{Enrico Daga}
\affiliation{%
  \institution{The Open University}
  \city{Milton Keynes}
  \country{United Kingdom}
}
\email{enrico.daga@open.ac.uk}

\author{Lucia Giagnolini}
\affiliation{%
  \institution{University of Bologna}
  \streetaddress{via Zamboni 32}
  \city{Bologna}
  \state{Italy}
  \country{Italy}
  \postcode{40126}
}
\email{lucia.giagnolini@studio.unibo.it}

\author{Francesca Tomasi}
\affiliation{%
  \institution{University of Bologna}
  \streetaddress{via Zamboni 32}
  \city{Bologna}
  \state{Italy}
  \country{Italy}
  \postcode{40126}
}
\email{francesca.tomasi@unibo.it}

\renewcommand{\shortauthors}{Daquino et al.}

\begin{abstract}
   Collaborative data collection initiatives are increasingly becoming pivotal to cultural institutions and scholars, to boost the population of born-digital archives. For over a decade, organisations have been leveraging Semantic Web technologies to design their workflows, ensure data quality, and a means for sharing and reusing (Linked Data). Crucially, scholarly projects that leverage cultural heritage data to collaboratively develop new resources would benefit from agile solutions to simplify the Linked Data production workflow via user-friendly interfaces. To date, only a few pioneers have abandoned legacy cataloguing and archiving systems to fully embrace the Linked Open Data (LOD) paradigm and manage their catalogues or research products through LOD-native management systems. In this article we present \emph{Crowdsourcing Linked Entities via web Form (CLEF)}, an agile LOD-native platform for collaborative data collection, peer-review, and publication. We detail design choices as motivated by two case studies, from the Cultural Heritage and scholarly domains respectively, and we discuss benefits of our solution in the light of prior works. In particular, the strong focus on user-friendly interfaces for producing FAIR data, the provenance-aware editorial process, and the integration with consolidated data management workflows, distinguish CLEF as a novel attempt to develop Linked Data platforms for cultural heritage.
\end{abstract}



\maketitle

\section{Introduction}\label{intro}
Collaborative data collection initiatives are increasingly becoming pivotal to cultural institutions and scholars, to boost the population of born-digital archives.
However, scholarly projects and applications in the cultural heritage domain have to comply with several requirements to ensure the FAIRness of their data \cite{wilkinson2016fair}, in order to support the reuse of such data and their long term availability. 
A number of scholarly data management workflows have been designed~\cite{lamprecht2020towards} to cope with issues related to data collection, update, and publication. 
To facilitate the task, user-friendly interfaces are required to collect and integrate data with external sources, to validate data quality, and to publish results that can be reused by a broad, diverse audience \cite{diulio2021usability}. 
In addition, storing provenance information along with content data is deemed essential to prevent inconsistencies when integrating sources, to emphasize content responsibility, and eventually to foster trust in data \cite{amin2008understanding,ceolin2016combining,sandusky2016computational}. Moreover, provenance is fundamental for project management purposes, e.g. to monitor the editorial process and to keep track of data versions. Finally, the usage of repositories for dissemination (e.g. GitHub\footnote{\url{https://github.com}}), and preservation (e.g. Zenodo\footnote{\url{http://zenodo.org/}}, Internet Archive\footnote{\url{https://archive.org}}) are suggested to improve findability, accessibility, and long-term availability of data. 

To cope with these requirements, cultural heritage institutions and scholars have been increasingly leveraging Semantic Web technologies. Consortia of museums, libraries, and archives \cite{doerr2010europeana,knoblock2017lessons,delmas2020fostering} adopt Linked Open Data (LOD) as a \emph{lingua franca} to develop data aggregators, promote crowdsourcing campaigns \cite{dijkshoorn2017accurator}, and serve high-quality data to scholars. Several solutions for provenance management have been discussed by the Semantic Web community \cite{heath2011linked,moreau2010foundations}, including the PROV Ontology \cite{lebo2013prov} and named graphs \cite{carroll2005named}, which have been largely evaluated in production environments \cite{sikos2020provenance,orlandi2021benchmarking}. In recent years, a few content management systems have been introduced to facilitate scholarly data publication and integration, such as Omeka S\footnote{\url{https://omeka.org/s/.}} or Semantic MediaWiki\footnote{\url{https://www.semantic-mediawiki.org/wiki/Semantic\_MediaWiki.}}, which store data in relational databases and provide means to serve Linked Data on demand. 


However, it has been argued that several issues affect Linked Data management and publishing systems themselves, spanning from storage to indexing and query-related issues \cite{hauswirth2017linked}. First, some solutions do not leverage provenance in data management workflows (e.g. Omeka S) and do not serve provenance data as Linked Open Data. In particular, provenance management in database approaches assumes a strict relational schema is in place, whereas RDF data is by definition schema free \cite{hauswirth2017linked}. Second, mechanisms to facilitate Linked Data collection and integration from external sources are not always implemented in user-friendly interfaces (e.g. Semantic MediaWiki), therefore time-consuming, error-prone, manual reconciliation tasks are still delegated to the user. Lastly, existing systems do not always offer tools for version control, continuous integration, and integration with consolidated data management workflows, which must be ensured by data providers separately.


As a matter of fact, only a few pioneers have abandoned legacy cataloguing and archiving systems to fully embrace the Semantic Web paradigm and manage their catalogues through LOD-native management systems \cite{malmsten2009exposing}. 
Institutions seem to prefer to maintain legacy systems for managing the data life-cycle (addressing aspects such as data entry, review, validation, and publication), and to provide dedicated services to access their 5-star data, whether these represent complete collections \cite{dijkshoorn2018rijksmuseum}, subsets \cite{deliot2014publishing,daquino2017enhancing}, or project-related data \cite{davis2019old}. 
In contrast to institutions, scholarly projects leverage cultural heritage Linked Data to develop new digital resources since the inception of their projects (see \cite{davis2021linked} for a recent survey).
Such projects often enrich descriptions of cultural heritage artefacts with novel contributions, and in turn they become precious sources of information for cataloguers, professionals, and scholars. A comprehensive Linked Open Data management environment would facilitate their tasks, such as data collection, data quality validation, and data publication, according to FAIR data requirements, so as to ensure seamless integration between diverse stakeholders' data sources.
While a few solutions have been developed by scholars to support Linked Data production with user-friendly, provenance-aware interfaces \cite{adamou2019crowdsourcing}, these solutions are not portable or reusable, and are difficult to maintain. This can hinder the use of such solutions, in particular for smaller institutions or projects that have limited technical support. 

 
In this article we introduce \emph{CLEF, Crowdsourcing Linked Entities via web Form}, an agile, portable, LOD-native platform for collaborative projects \cite{marilena_daquino_2021_5603223}. The objective of CLEF is threefold, namely: to facilitate Linked Open Data creation, integration, and publication using user-friendly solutions; to leverage provenance information stored in named graphs in data management and editorial workflows; and to integrate Linked Data production with existing data management workflows. CLEF has been designed with a bottom-up approach, taking into account requirements highlighted by two case studies, respectively representative of (i) crowdsourcing campaigns promoted by cultural institutions and (ii) collaborative data collections in scholarly projects. 
Unlike most existing solutions, CLEF manages Linked Open Data from the beginning of a project, provides means to build user interfaces easily, and manages provenance information as first-class entities. In particular, CLEF is integrated with GitHub, which allows fine-grained provenance information to be traced and to be harmonised with common data management workflows. 

The remainder of the article is as follows. In section \ref{background} we present the two exemplary case studies motivating the design of CLEF. In section \ref{lodnative} we introduce CLEF's requirements, architecture overview, provenance management, data integration features, and sustainability model. In section \ref{related} we review existing solutions for Linked Data creation and publication, highlighting gaps and opportunities with respect to the requirements we identified in the case studies.
In section \ref{discussion} we discuss benefits, opportunities and limitations derived from the development of CLEF in the light of existing solutions. Finally, in section \ref{conclusion} we conclude with future work.
   


\section{Background and case studies}\label{background}

The landscape of collaborative projects in the Cultural Heritage domain is broad. Several participation models are in place, such as crowdsourcing campaigns promoted by cultural institutions and companies, scholarly collaborative projects, platforms for peer-reviewing content, and so on \cite{daquino2021linked}. Campaigns may require limited contributions from users, like \emph{social tagging}, or can require transcription of texts and correction of metadata. Projects may collect and host contents provided by users, like stories about cultural objects preserved by the institution, or may accept new objects and descriptions curated by users. 

Moreover, crowdsourcing campaigns may target a specific community or can be a short-term activity, focused on a particular cultural heritage collection. Investing in updating a legacy data management solution to fit the needs of a focused activity would be overkill. Often, such systems are developed \emph{ad hoc} (e.g. Google Forms) and can only satisfy the data collection requirement. In contrast, existing full-fledged data management solution would necessarily require major changes for satisfying the needs of a local project (further discussed in section \ref{related}). 

In this work we consider two common scenarios, namely: (i) a crowdsourcing campaign promoted by cultural institutions to aggregate metadata and produce new information sources for qualitative and quantitative analysis (section \ref{ARTchives}), and (ii) a scholarly project to collaboratively collect information on digital heritage resources and provide an online catalogue for discovery purposes (section \ref{musoW}). We collected requirements from two real-world scenarios, respectively the \emph{ARTchives} and \emph{musoW} projects, during interviews with stakeholders (ARTchives, musoW) and project meetings (musoW). The added value of such type of projects to the economy of cultural heritage data is of great importance for institutions, which foster co-curation practices as a way to engage with patrons, to disseminate high quality contents, and to promote cultural awareness and, partly, participation \cite{daga2021integrating,bonacchi2019participation}.
In the following sections we present the case studies, highlighting the categorization of user requirements in parentheses.

\subsection{ARTchives: a crowdsourcing campaign for art historians' archival collections}\label{ARTchives}


ARTchives\footnote{\url{http://artchives.fondazionezeri.unibo.it/}} is a collaborative project supported by six institutions (Bibliotheca Hertziana, Rome;
Federico Zeri Foundation, Bologna;
Getty Research Institute, Los Angeles;
Kunsthistorisches Institut in Florenz, Florence;
Scuola Normale Superiore, Pisa;
Università Roma Tre, Rome) which aims at surveying the heritage of art historians' archival collections \cite{daquino2021artchives}. The objective of ARTchives is to provide scholars with an online tool to retrieve information about archival collections relevant to their studies, gather bibliographic sources, and answer research questions related to Art historiography with quantitative methods - such as historians’ network analysis, topic analysis of debates, interlinking collections etc. 

Since a wealth of information is already available on the Web, the project aims at referencing existing data sources as much as possible and creating new data only when needed (\textbf{reusability}). Existing data include bibliographic references from the Open Library\footnote{\url{https://openlibrary.org/}}, historical information from Wikidata\footnote{\url{http://wikidata.org/}}, technical terms from the Getty Art and Architecture Thesaurus\footnote{\url{https://www.getty.edu/research/tools/vocabularies/aat/}}, and biographical information from the Dictionary of Art Historians\footnote{\url{https://arthistorians.info/}}. In particular, a few sources (e.g. biographies) are available as long natural language texts, which cataloguers would copy-paste, and which would require knowledge extraction mechanisms to identify important entities and relations (\textbf{enhancement}). Cataloguers must be able to accept or reject automatic suggestions (\textbf{accuracy}). The latter can be considered the main contribution of ARTchives, which allows users to build bridges between well-known data sources in Art history, thus simplifying interlinking operations. Moreover, the description of archival collections complies with archival metadata standards, for which existing ontologies shall be reused, and must conform to user-specified templates (\textbf{validation}). 

Contributors in ARTchives are historians, cataloguers, and archivists belonging to cultural institutions preserving art historians' collections. Members of an internal editorial board review data entered by guest cataloguers before publication. Since several institutions may contribute collections from the same art historian, thereby providing contrasting information on the same people, the peer-review process plays an important role in ensuring data quality and consistency (\textbf{consistency}). Existing platforms do not allow keeping track of competing contributions to the same artefact, which are usually incompatible, and only one version (the latest) is stored. In ARTchives, cataloguers would need to prevent data duplication, e.g. by informing cataloguers of existing duplicates, and allowing competing information to be temporarily stored in separate records. Reviewers could validate the one to be shown in the final application (\textbf{accuracy}).

Already at the inception of the project it was clear that the sustainability of data created by ARTchives would be hampered in the long run, since the maintenance of the project is affected by time and resource constraints. To this extent, ARTchives is representative of many small-medium research projects that struggle to maintain their infrastructure in the long run (\textbf{preservation}) and would need to donate their data to other projects or initiatives. 


\subsection{musoW: a collaborative catalogue of music resources on the Web}\label{musoW} 

Musical data play a key role in the every-day life of musicologists, music teachers/learners, and creative industries, who often need to combine diverse resources (e.g. music scores, audiovisual materials, data) from digital music libraries and audiovisual archives for their purposes. In musoW\footnote{\url{https://w3id.org/musow}} stakeholders can gather information on musical resources available on the Web and incrementally populate an online catalogue to be used for discovery purposes, e.g. to collect online sources relevant to musicology enquiries or music teaching \cite{daquino2017characterizing}.
Records in musoW include descriptions of online resources, such as digital libraries, repositories, datasets, and relevant software solutions. For each resource bibliographic data, scope insights, and technical information are recorded, e.g. responsible people or projects, relevant musicians or music genres, data licenses, APIs and other services. 
Unlike ARTchives, also non-experts can contribute to musoW, who may not be familiar with terminology and existing records in the musoW catalogue. In order to facilitate interlinking records and data consistency, auto-completion suggestions should be provided from controlled vocabularies and existing records (\textbf{reusability}). 

musoW is part of Polifonia, an EU-funded project which aims to interlink musical heritage resources and produce tools to effectively support scholars, professionals, and people passionate about music in knowledge discovery. Members and accredited contributors of the Polifonia organization repository\footnote{\url{https://github.com/polifonia-project/}} can validate crowdsourced records and publish them (\textbf{accuracy}). To maximise data reuse, data should be easily accessible and securely backed up (\textbf{accessibility}). To track fine-grained provenance of records (\textbf{reusability}), every change occurring in records should be registered. 

Moreover, to ensure services built on top of musoW can rely on continuous data availability, once a record is published it cannot be unpublished. Rather, records must be flagged as \emph{drafts}, so that it is clear to both users and applications which publication stage the data is at (\textbf{persistency}). 

musoW is meant to increase the findability of collected resources, which would be otherwise only known to hyper-specialised communities (\textbf{findability}). To do so, indexing of resources in well-known search engines is deemed a priority (\textbf{discoverability}).

Moreover, to simplify data exploration and discovery, browsing solutions including indexes, filters, and aggregations of records should be generated according to preferences specified in record templates (\textbf{exploration}). Since such preferences may change over time along with the population of new data, the definition of new filters and data views should be easy to do and have immediate effect. Lastly, part of the project mission is to contribute to the long term preservation of catalogued resources (\textbf{preservation}).

\section{Linked Open Data native cataloguing with CLEF}\label{lodnative}

To design CLEF we started from the user requirements collected from the aforementioned scenarios in a bottom-up fashion.
We realised that most user requirements highlighted in the case studies correspond to well-known requirements in Linked Data publishing practices \cite{world2014best}, as well as challenges in the development of Linked Data Platforms, as specified (in a top-down approach) by prior works \cite{mihindukulasooriya2013linked}. Our research interest in developing CLEF is to test the strength of such parallelism, and understand the benefits and limitations of LOD-native approaches for solving common problems in the Cultural Heritage domain. Therefore, where applicable, we addressed requirements as design issues of a Linked Open Data management workflow. 

\subsection{Requirements}\label{requirements}



For the sake of readability, we grouped requirements into four categories, previously acknowledged as FAIR principles (Findability, Accessibility, Interoperability, Reusability) \cite{wilkinson2016fair}. In so doing we want to stress one of the main strengths of CLEF, i.e. producing high-quality, reusable data. This aspect is particularly relevant to the scholarly domain, and in recent years it has been increasingly addressed in the Cultural Heritage domain too \cite{hermon2021fair}.
Along with requirements, we propose one or more solutions to drive development. Where applicable, such requirements have been translated into Linked Open Data requirements or functionalities. 

\begin{itemize}
    \item \textbf{Findability}. Data are identified with persistent identifiers (URIs), described with rich metadata, and are findable in the Web.
    \begin{itemize}
        \item \emph{Discoverability. Allow search engines to leverage structured data for indexing purposes.} Every record is served as HTML5 documents including RDFa annotations.
        \item \emph{Exploration. Automatically generate data views to facilitate retrieval.} Operations for automatically generating views, such as filtering, grouping, and sorting, are available. These are ontology-driven, i.e. the result of SPARQL queries.
    \end{itemize}
    \item \textbf{Accessibility}. Data are accessible via the HTTP protocol, and are available in the long term via a plethora of solutions for programmatic data access.
        \begin{itemize}
            \item \emph{Preservation (sources). Request digital preservation of user-specified resources.} Rely on established services like the WayBack machine\footnote{\url{https://archive.org/web/}} for web archiving. 
            
            \emph{Preservation (ontologies). Allow direct reuse of up-to-date schemas and data of existing projects.} Retrieve information on the user-defined data model from the Linked Open Vocabularies initiative \cite{vandenbussche2017linked}.
            
            \emph{Preservation (data). Integrate the system with established data management workflows.} Bind changes in Linked Data to commits in GitHub and release versions in Zenodo.
            
            \item \emph{Persistency. Ensure continuity of services built on top of generated data.} Prevent deletion of published records identified with persistent URIs.
        \end{itemize}
    \item \textbf{Interoperability}. Data are served in standard serialisations, include references to standard or popular ontologies, and links to external Linked Open Data sources. While this was not an explicit requirement highlighted from use cases, it is a natural consequence of the usage of Linked Open Data, which makes it easier to work with data (e.g. in data integration over multiple sources). 
    
    \item \textbf{Reusability}. Data are released as open data with non-restrictive licenses, are associated with detailed provenance information and follow well-known data sharing policies.
        \begin{itemize}
            \item \emph{Enhancement. Generate structured data from natural language texts.} Perform Named Entity Recognition over long texts on demand, extract structured data, and reconcile to Wikidata. 
            \item \emph{Consistency. Ensure  interlinking of records and correct usage of terminology.} Suggest terms from selected Linked Open Data sources and user-specified controlled vocabularies while creating new records. Allow contradictory information to be recorded as named graphs. Ensure peer-review mechanisms are enabled to supervise contributions from non-experts, and prevent inconsistencies in the final user application. Allow restriction of access, and give privileges to a group of users that share ownership of data on GitHub.
            \item \emph{Accuracy. Allow fine-grained curatorial intervention on crowdsourced data.} Represent records as named graphs and annotate graphs with provenance information according to the PROV ontology (including contributors, dates, and activities/stages in the peer-review process). Update annotations every time a change happens in the graphs. Track changes and responsibilities in GitHub commits.
            \item \emph{Validation. Allow automatic validation of data.} Along with manual curation, perform schema and instance level checks to ensure created data conform to user-generated (ontology-based) templates.
        \end{itemize}
     
\end{itemize}

Moreover, while not in scope in FAIR principles, the use cases highlighted that in order to prevent error-prone operations, guarantee high-level data quality standards, and serve easy-to-find data, \textbf{user-friendly interfaces} are necessary or highly recommended. Therefore, the provision of easy-to-use interfaces becomes a fundamental user requirement of CLEF to ensure (1) reusability of data and easiness of exploration for the final user, and (2) simplicity and error avoidance for editors and administrators.

In summary, the interaction with stakeholders highlighted three important research areas, namely: (1) the need of user interfaces to manage most data management processes, which would otherwise require complex or time-consuming operations to be performed manually (e.g. data reconciliation, data quality validation, data exploration) (UF); (2) the importance of provenance management in the editorial process (PM); and (3) the compliance with reusability and sustainability requirements and the integration with data management workflows for scholarly data (DMI). Managing data natively as Linked Data allows us to address all three aspects and to fully comply with FAIR principles.

\subsection{CLEF overview}\label{architecture}

CLEF is a highly configurable application that allows digital humanists and domain experts to build their own crowdsourcing platform, to integrate the data management workflow with Linked Open Data standards and popular development and community platforms, and immediately enjoy high-quality data with exploratory tools. CLEF is a web-based application in which users can describe resources (e.g. real-world entities, concepts, digital resources) via intuitive web forms. To help with entering descriptions, users are offered auto-complete suggestions from vocabularies, existing records and terms automatically extracted from text. 
Administrative users have full control of the setup of their CLEF application and of the definition of templates for describing information about their resources. The templating system of CLEF is configurable via a web interface, in which each form field for describing the resource is mapped to an ontology predicate chosen by the user. Templates are the main drivers of the application, since they ensure consistency in data entry and data validation, they guide the peer-review of records, and are fundamental in retrieval and exploration via actionable filters. It's worth noting that the template setup, in which the ontology mapping is manually (at present) curated, is the only input that requires expert users. 

Both authenticated and anonymous contributions can be enabled. A simple peer-review mechanism allows users to curate their records, publish them, and continuously populate the catalogue of contents, which can be immediately browsed via automatically generated interfaces and filters. The tool is particularly suitable for collaborative projects that need to restrict access to members of one or more organisations, to share data and code on dissemination repositories, and that need an environment to discuss project issues. In fact, CLEF is designed to be easily integrated with GitHub and simplify the data management workflow, naturally supporting several of the FAIR requirements.

\begin{figure}[htp]
    \centering
    \includegraphics[width=12cm]{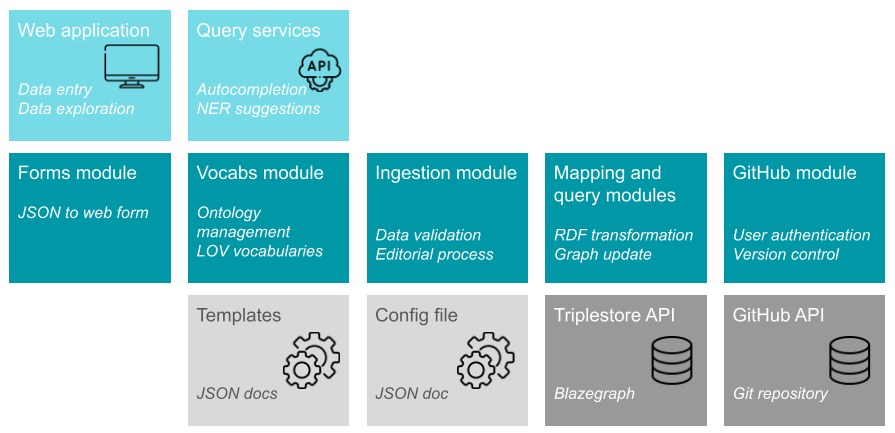}
    \caption{CLEF overview}
    \label{fig:clef}
\end{figure}

Fig. \ref{fig:clef} presents an overview of the CLEF data management system. In detail, CLEF allows an administrator to configure and customise the application via user-friendly interfaces. 
In the \textbf{configuration} setup, users can specify information relevant to their dataset, e.g. URI base, prefix, SPARQL endpoint API (with default configuration), and optional mechanisms for version control and user authentication (via GitHub). The setup of dereferencing mechanisms is delegated to the adopter, who can choose and set up redirection rules by means of their favourite persistent URI provider (e.g. w3id\footnote{\url{https://w3id.org/}}). 

For each type of resources to be collected and described, a \textbf{template} is created in the form of a JSON mapping document. This includes form field types (e.g. text box, checkbox, dropdown), expected values (literals, entities), services to be called (e.g. autocomplete based on Wikidata and the catalogue, Named Entity Recognition in long texts), the mapping between fields and ontology terms or controlled vocabularies, and whether the field should be used in a default web page, called \emph{Explore}, as a filter to aggregate data. 

Ontology terms and terms from controlled vocabularies specified by users are managed via the \textbf{vocabulary} module. It's worth noting that, while users can specify their own ontology terms, CLEF fosters reuse of popular and standard vocabularies. The module updates the CLEF triplestore with user-specified terms (which may be new terms or terms belonging to existing ontologies) and calls the APIs of \emph{LOV Linked Open Vocabularies} \cite{vandenbussche2017linked} to retrieve original labels and comments associated with reused terms. The resulting data model is shown in a dedicated web page called \emph{Data Model} along with information retrieved from LOV.  
    
The \textbf{form} for data entry is generated according to settings specified in templates. While editing (creating, modifying, or reviewing) a record, both CLEF triplestore and external \textbf{services} like DBPedia spotlight\footnote{\url{https://www.dbpedia-spotlight.org/}} and Wikidata APIs are called to provide suggestions. Every time a record is created/modified, data are sent to the \textbf{ingestion} module. The latter performs a first validation of the form based on the associated templates, and calls the \textbf{mapping} module, which transforms data into RDF according to ontology terms specified in the template and updates the named graph created for the record. 

CLEF supports a compliant SPARQL 1.1~\cite{sparql11} endpoint as back-end, therefore, it is not dependent on a specific implementation.
However, current running instances use Blazegraph~\cite{thompson2016bigdata}. 
In particular, named graphs are extensively used to annotate and retrieve provenance information needed to manage the peer-review process, and to efficiently serve record-related information in the exploratory interfaces.

A module is dedicated to the interaction with \textbf{GitHub}. GitHub was chosen for its popularity as a dissemination platform for versioned code and data, which fosters visibility of project results, and for its services - i.e., APIs for read-write operations, OAuth mechanisms. Users may decide to bind their application to a GitHub repository, which allows them (1) to store a backup of data in a public/private repository, (2) to keep track of every change to data via commits, and (3) to enable user authentication to the web application via GitHub OAuth\footnote{\url{https://docs.github.com/en/developers/apps/building-oauth-apps/authorizing-oauth-apps}}.  

Lastly, to increase findability of collections, a few automatically generated web pages serve browsing and search interfaces over the catalogue. Currently CLEF provides the following templates: a homepage; the backend controller from which to access the list of records, the setup configuration form and the templates forms; records creation/modification/review and publication forms; a \emph{Documentation} page with instructions on the usage of forms; the \emph{Explore} page, where views on collected data are shown and filtered; a template to display the records wherein Linked Open Data are also served as RDFa annotations; a template to display controlled vocabulary terms and statistics on their usage in the catalogue; a \emph{Data model} page, collecting ontology terms and definitions from LOV; and a GUI to query the SPARQL endpoint.

The software has been developed in two phases.
An initial data management system was developed for ARTchives\footnote{ARTchives source code is available at: \url{https://github.com/marilenadaquino/ARTchives}}. In a second phase, the code base has been extended and adapted to be customizable and reusable as-is in other crowdsourcing projects. CLEF is developed in Python, based on Webpy\footnote{\url{https://webpy.org}}, a simple and small-size framework for web applications. The source code of CLEF is available on GitHub\footnote{CLEF source code available at: \url{https://github.com/polifonia-project/clef/releases/latest}, ISC license.} and Zenodo \cite{marilena_daquino_2021_5603223}. 

\color{red}
CLEF is a production-ready solution. It is under continuous development to become a flexible tool for a wider range of collaborative scholarly projects. Potential scalability issues have been improved in recent SPARQL / quad store implementations and there is a continuing effort in the community to support analysis of performance \cite{herrera2019btc,roder2020hobbit}.
\color{black}
The initial version of the system was tested with around 15 cataloguers of the six institutions promoting the ARTchives project, namely: the Federico Zeri Foundation (Bologna), Bibliotheca Hertziana (Rome), Getty Research Institute (Los Angeles), Kunsthistorisches Institut in Florenz (Florence), 
Scuola Normale Superiore (Pisa), and 
Università Roma Tre (Rome). Currently, user tests are continuously performed by Polifonia project members, who provide new requirements to foster development and research, documented in the musoW repository issue tracker\footnote{\url{https://github.com/polifonia-project/registry_app/issues}}. User tests will soon be performed with users with different profiles and less technical experience.

\subsection{The editorial process: provenance management and user authentication}\label{editorial}

In CLEF, every record is formally represented as a named graph \cite{carroll2005named}. Named graphs enable us to add RDF statements to describe those graphs, including their provenance, such as activities, dates, and agents involved in the creation and modification of a record. Provenance information is described by means of the well-known W3C-endorsed PROV Ontology \cite{lebo2013prov}. 
Moreover, named graphs allow us to prevent inconsistency caused by competing descriptions for the same entities, for instance when different cataloguers describe the same creator of multiple collections. While this scenario is allowed, users are informed of existing potential duplicates when creating a new record, which prevents involuntary inconsistencies. 


The editorial process in CLEF addresses three phases: record creation, record modification, and review and publication. When creating a record, the corresponding named graph is annotated with the identifier of the responsible user (anonymous user if no authentication method is set), the timestamp, and the publication stage \texttt{unmodified}. When modifying a record, additional provenance information is added, including the identifier of the (new) responsible user, the new timestamp, and the new stage (\texttt{modified}). Lastly, when publishing the record, the stage changes to \texttt{published}. A published record can be browsed from the \emph{Explore} page, searched from a text search, and can be retrieved as Linked Open Data from the SPARQL endpoint, via the REST API at \texttt{<APP-URL>/sparql}. While a published record can be modified, and therefore moved back to the stage \texttt{modified}, it cannot be unpublished. While this may be inconvenient in some scenarios, this prevents applications relying on records (and related persistent URIs) from getting unexpected, inconsistent responses.

We chose GitHub to manage user authentication, fine-grained provenance tracking, and version control.
In general, CLEF allows both authenticated and anonymous users to create new records. However, records can be modified and published only by accredited users. CLEF is optimised to authenticate users that have a GitHub account. To enable GitHub authentication in the initial setup of CLEF, users must specify (1) their GitHub credentials, (2) a GitHub repository they own, and (3) must have created an OAuth App connected to their repository, so as to enable read-write operations on the repository and to confirm the identity of collaborators. 

Every time a change is made to a record, content data and provenance information are updated on the triplestore via its REST API, on the file system, and - if enabled - also on GitHub. To avoid spamming, only records that have been reviewed are stored on GitHub, thereby initialising the versioning. All changes to records are identified by a commit on the repository, and it is possible to track which information (i.e. field of the resource template) has been modified. While such information is currently not stored as Linked Open Data, auxiliary tools such as git2PROV \cite{de2013git2prov} can be used to generate PROV-compliant RDF data. 
In so doing, we prevent the development from scratch of features that are anyway available on Github, and we intertwine the two platforms for a better data management workflow. 

In case a user decides not to enable the GitHub synchronization, data are stored in the local triplestore, changes in data are recorded with minimal provenance information (date of changes and publication stage), and only anonymous contributions can be made to the platform. The latter scenario is particularly handy if the application runs only locally (e.g. because contributors do not have the possibility to run the application on a remote server). Indeed, users may decide to create data via their own private instance of CLEF (which runs as a web application in localhost), to store data on their local triplestore, and to manage the publication as they prefer.  Moreover, if the application runs locally and user authentication is not enabled, but local users have a GitHub account and collaborate on a GitHub repository with other users, they may decide to keep working locally with CLEF, and backup their data on the shared repository. While publishing a remote instance of CLEF without any user authentication method is discouraged, CLEF implements anti-spamming mechanisms, in order to limit contributions from IP addresses, and to disable write operations on the triplestore. 

\subsection{Support data collection: reconciliation and enhancement}\label{wikidata}

When creating or modifying a record, contributors are supported in certain tasks relevant to the reusability of their data, namely: (1) data reconciliation, (2) duplicate avoidance, (3) keyword extraction, (4) data integration. 

In detail, when field values address real-world entities or concepts that can appear in other records, autocomplete suggestions are provided by querying  external selected sources (live) and the SPARQL endpoint of the project at hand. Suggestions appear in the form of lists of terms, each term including a label, a short description (to disambiguate homonyms) and a link to the online record (e.g. the web page of Wikidata entity or a record already described in the project). If no matches are found, users can create new entities that are added to the knowledge base of their project and these will appear in the list of suggestions in new records. Currently CLEF is optimised to work with Wikidata, but implementations of entity linking from the Open Library, the Getty AAT, and the Getty ULAN are available too.

When designing a resource template, users can flag a specific field to be used for disambiguation purposes (e.g. the field \emph{title} for a book, the field \emph{name} for a person). When creating a new record, the specified field is bound to a lookup service that alerts the user of potential duplicates already existing in the catalogue. The user may accept or ignore the recommendation.

Some fields may require contributors to enter long free-text descriptions (e.g. historians' biographies, scope and content of collections), which include a wealth of information that cannot be processed as machine-readable data. To prevent such a loss, two concurrent Named Entity Recognition (NER) tools (i.e. DBpedia spotlight API\footnote{\url{https://www.dbpedia-spotlight.org/api}} and compromise.js\footnote{\url{http://compromise.cool/}}) extract entity names (e.g. people, places, subjects) from the text. Extracted entities are reconciled to Wikidata entities and keywords (bound to Wikidata Q-IDs) and are shown to users to approve/discard. Approved terms are included in the data as machine-readable keywords associated to the subject entity. 

When Wikidata terms are reused, the system can be configured to query the Wikidata SPARQL endpoint to retrieve and store context information in the knowledge base. For instance, in ARTchives, Wikidata entities representing artists, artworks, and artistic periods (recorded as subjects addressed by contents of archival collections) are automatically enriched with time spans, retrieved from the Wikidata SPARQL endpoint and saved in the local triplestore; likewise, Wikidata entities representing historians are enriched with birth and death places. Finally, entities can be geo-localised via OpenStreetMap APIs\footnote{\url{https://www.openstreetmap.org/}}.

\subsection{Data sustainability: ontologies, data, and long-term preservation strategies}\label{modelling}

Long-term accessibility of scholarly projects is often hampered by time and resource constraints.
A well-known problem is the maintenance of ontologies adopted by small-medium crowdsourcing or scholarly projects \cite{carriero2020landscape}. While CLEF does not prevent the creation of new ontology terms, which are stored along with data, CLEF supports reuse of external ontologies. Terms from external ontologies can be directly referenced in resource templates to map form fields to predicates and to map templates themselves to classes. Where reused ontologies are popular or W3-endorsed ontologies, CLEF allows enriching referenced terms with definitions provided by Linked Open Vocabularies (LOV). Note that reused ontologies are not imported. This design choice has the evident drawback of preventing inference mechanisms, which are not applicable without manually importing ontologies in the knowledge base created by CLEF. Nonetheless, due to this design choice, we believe CLEF has the merit of complying with another debated requirements in the Semantic web community \cite{carriero2020landscape}, namely, the ability to rely on up-to-date information on reused ontologies, provided by LOV. 

Like the projects themselves, the wealth of data produced by scholarly initiatives often becomes unavailable in the mid/long-term. To prevent that, CLEF adopts several strategies. First, CLEF is optimised to reuse Wikidata as much as possible, both at schema level (users can choose classes and properties from Wikidata data model) and at instance level (autocompletion suggestions reuse individuals from Wikidata). The idea is to support stakeholders in producing curated metadata that can can be exported and imported into Wikidata according to its guidelines for contributors\footnote{\url{https://www.wikidata.org/wiki/Wikidata:Data\_donation}}. While Wikidata allows users to also import non-Linked Data into the knowledge base, and to manually perform entity matching, CLEF data include entities already matched with Wikidata Q-IDs, avoiding the need for manual matching. Data can be retrieved via the SPARQL endpoint or via the GitHub repository. 

Second, by synchronising CLEF knowledge graphs with GitHub it is also possible to synchronise the repository with Zenodo. Zenodo is a certified repository for long-term preservation, widely recognized in the scientific community. Zenodo has recently offered the opportunity to link GitHub repositories to their platform, binding GitHub releases to new versions on Zenodo, uniquely identified with a DOI.   

Lastly, the case studies highlighted the need to access and extract information from online web pages (e.g. the Dictionary of Art Historians, online music resources) and reference the source in records. Such web pages are cited as sources of information or are described as first-class entities in records, and are likely to be explored by final users of the project catalogues. Ensuring the persistence of such pages in the long-term is an important aspect, which contributes to foster trust in scholarly and cultural heritage projects. While preserving the original web sources along with data created in CLEF would be inconvenient for small-medium projects - that cannot afford to archive all the web sources they mention - CLEF allows users to specify which form fields include URLs that should be sent to the Wayback machine\footnote{\url{https://web.archive.org/web}}, which in turn takes a snapshot of the webpage and preserves it. 

\section{Related work}\label{related}

Over the years special purpose systems have been designed by cultural heritage institutions to systematically collect user-generated data and serve Linked Open Data. However, such software and initiatives do not address all our three main requirements - i.e. user-friendliness for final users (UF-U) and administrators (UF-A), provenance management (PM), and data management integration (DMI). Moreover, researchers have argued that solutions
developed for a single institution or project turn out not to be sustainable or not motivating for users \cite{koukopoulos2017trustworthy}. Therefore we also consider sustainability and reusability out of the original context (SR) when reviewing prior works. An overview of surveyed systems is shown in table \ref{survey}.

\begin{center}
\begin{table}[h!]
\begin{tabular}{ | m{9em} | m{2cm} | m{2cm} | m{2cm} | m{2cm} | m{2cm} | } 
  \hline
  Name & User friendly (Users) & User friendly (Admin) & Provenance Mgmt. & Data Mgmt. integration & Sustainability Reusability \\ 
  \hline
  LED & \checkmark & \checkmark & \checkmark & \checkmark &   \\ 
  \hline
  OmekaS & \checkmark & \checkmark &  &  & \checkmark  \\ 
  \hline
  Semantic MediaWiki & \checkmark & \checkmark & \checkmark &  & \checkmark  \\ 
  \hline
  Sinopia & \checkmark & \checkmark &  &  & \checkmark  \\ 
  \hline
  ResearchSpace & \checkmark & & \checkmark & \checkmark & \checkmark  \\ 
  \hline
\end{tabular}
\caption{Overview of systems for collaborative data collection and Linked Data publishing}\label{survey}
\vspace{-10mm}
\end{table}
\end{center}


Among the scholarly projects that have been leveraging Semantic Web since early stages of data collection, we acknowledge the Listening Experience Database (LED) \cite{adamou2019crowdsourcing}, which exceptionally adopts Semantic Web technologies to support the entire life-cycle of data management, from data collection to publication, by means of user-friendly interfaces (UF-U, UF-A). It offers user-friendly interfaces for data entry, peer-review, and exploration, leveraging several external services and sources, such as the British National Bibliography, DBpedia, and LinkedBrainz. Moreover, each record contributed by users is represented as a named graph, and provenance is accurately annotated for each graph (PM). Data are served via a SPARQL endpoint and a daily backup is provided as a link on the website (DMI). 
Currently, LED relies on an ad-hoc application developed to serve project-related goals. The software relies on a heavily customized version of the Drupal CMS, which is not adaptable to support different data models and ontologies and therefore it cannot be of immediate reuse in projects with a different scope (SR).

In recent years, a few content management systems have been introduced to facilitate new projects to publish Linked Data via reusable platforms. 
Omeka S\footnote{\url{https://omeka.org/s/.}} is a popular platform for collaborative data collection and creation of virtual exhibitions (UF-U, UF-A). User groups and roles can be defined. However editors do not have the means to supervise the peer-review process on records, including important provenance-related aspects like changes made by contributors in records, flagging records as under review or ready for publication, which must be manually included by users (e.g. records form fields) or managed separately (PM). Records (also called \emph{items}) are served as JSON-LD documents via API. However, data cannot be accessed in any other RDF serialisation and cannot be queried via a SPARQL endpoint, which may be cumbersome in some situations, e.g. the exploration of the dataset for analysis purposes. CSV data exports can be requested on demand via a dedicated plugin. However, no mechanism for automatic data versioning is active (DMI). It is also worth noting that the visibility (i.e. the availability) of records can be constantly modified, thereby hindering the continuity and reliability of services relying on data served by the application (DMI). The software is open source, actively developed and maintained by a broad community and serves several projects in the Cultural Heritage domain (SR).

Semantic MediaWiki\footnote{\url{https://www.semantic-mediawiki.org/wiki/Semantic\_MediaWiki.}} is another popular tool used in well-known projects like Wikipedia, which allows data to be displayed in a catalogue fashion (UF-U). New records can be created via web forms and external sources can be used to populate fields when dedicated plugins are installed (UF-A). The system enables fine-grained editorial control (PM) and serves data as LOD, which are stored in a triplestore on which SPARQL queries can be performed (DMI). Like Omeka S, Semantic MediaWiki is open source, actively maintained, and supported by a broad community (SR). 



Sinopia\footnote{\url{https://sinopia.io}} is a web-based environment developed by the LD4P (Linked Data for Production) initiative, based on Library of Congress’s BIBFRAME Editor and Profile Editor\footnote{\url{https://bibframe.org}}. Sinopia potentially supports other ontologies than BIBFRAME, and users can customize templates for the creation of RDF triples via web forms (UF-A). Unfortunately, Sinopia does not provide a fully-fledged editorial workflow. In fact, provenance information must be recorded manually by cataloguers, which is otherwise not stored along with data (PM). Like in OmekaS, data are stored in a relational database and are shared via APIs that serve JSON-LD documents (DMI). The software is currently under development (SR).


ResearchSpace\footnote{\url{https://https://researchspace.org/}} is a semantic web platform that allows heritage institutions to create and publish collections as LOD. Indeed, the platform is optimised to build exhibitions (UF-U), and it allows the creation of sophisticated browsing interfaces via user templates. While providing flexibility and freedom to customise templates according to user-generated Linked Data patterns, unfortunately, an expert user is needed to encode such preferences in the template, using a combination of HTML, SPARQL, Javascript, and a custom templating language (UF-A). Likewise, the entire setup of the back-end functionalities requires such a templating system to be manually setup (UF-A). Notably, when creating new records, provenance information is stored along with the data (PM). Linked Data are stored in a triplestore and a SPARQL endpoint REST API is provided. Assets like ontologies and vocabularies can be versioned by configuring a Git repository (DMI). Data can be uploaded to the triplestore via the back-end, and can be manipulated with several types of visual authoring tools (e.g. image annotators, PDF annotators, graphical network editor) (UF-U). 
The software is based on Metaphactory \cite{haase2019metaphactory}, a commercial enterprise product, customised to comply with the requirements of museums and libraries supporting ResearchSpace (SR).

\section{Discussion}\label{discussion}

We collected requirements to develop CLEF from two paradigmatic case studies, which address collaborations between scholars and cultural heritage institutions, an increasingly common scenario in the landscape of Linked Data and Cultural Heritage \cite{davis2021linked}.

Such requirements have previously been tackled by other projects or software solutions. Each of these projects address certain of the requirements, but not all. 
In this section we first discuss strengths and shortcomings of prior works, and then we show how CLEF addresses these problems. Secondly, we highlight - where applicable - the benefits and pitfalls of using Linked Open Data technologies for this task. 

\emph{User-friendly ways to build interfaces.} Among surveyed systems, ResearchSpace offers the most powerful, flexible templating system and engaging visual aids for data manipulation and annotation. However, it requires expert users to set up all the browsing interfaces by means of a custom markup language. This aspect may negatively affect organisations that do not have specialised personnel available for the entire duration of the project. 
In CLEF we minimize the need of technical expertise to the specification of ontology terms underneath the templating system, which assumes the administrator has basic knowledge of ontology specifications, and the optional customisation of browsing templates. All browsing templates are provided with a default HTML configuration, which allows users to benefit from CLEF exploratory interfaces from day zero, without requiring any technical expertise. Moreover, the HTML templating system used by CLEF is implemented in pure Python+HTML. Therefore, in case users need to modify the functionalities or the look and feel of the web pages, a programmer would not require any particular knowledge of custom (enterprise) languages. 


\emph{Provenance management.} Despite being a popular solution, Omeka S does not shine in guaranteeing consistency, accuracy, and continuity of data produced. The lack of editorial control and the flexibility in hiding the visibility of resources identified with persistent URIs are two important shortcomings that affect organisational workflows and reliability of data sources. Likewise, Sinopia does not provide built-in methods to track provenance and support editors in guaranteeing accuracy of records. On the other hand, LED provides fine-grained provenance description, which is stored and queried along with content data, and reuses state-of-the-art Semantic web standards, such as the PROV ontology and RDF named graphs.
LED was an inspiration to develop the provenance management strategy for CLEF. Moreover, the integration of CLEF with GitHub allowed us to move a step forward towards an even more detailed provenance tracking mechanism. The download of records as RDF/ttl files in a Git repository, and the commit of every change done to records modified via the web application, enables a refined editorial workflow, in which expert users can grasp changes made on individual triples/form fields.   

\emph{Data management integration.} Considering the sustainability of data produced by projects, we observed that none of the prior solutions really tackles data management workflows by taking into account well-known good practices. ResearchSpace allows versioning of static assets like ontologies and vocabularies, but does not provide versioning mechanisms for data. LED adopts versioning mechanisms and provides a dump of data on a daily base, but these are not citable via a DOI. Since no description is provided, we assume such mechanisms are not integrated with the platform for data collection, therefore requiring extra effort in developing dissemination and preservation strategies. OmekaS allows data exports, but no mechanism for creating releases is active.
As a matter of fact, most projects reusing such software solutions do not provide versioned, citable, open data\footnote{We reviewed the list of projects adopting ResearchSpace (\url{https://researchspace.org/projects-and-collaborations/}) and Omeka S (\url{https://omeka.org/s/directory/}), which revealed that the majority of research products do not release (and share) their data via any dissemination or long-term preservation platform, such as GitHub and Zenodo.}. 
CLEF solves this problem by integrating with common data management workflows. Users can do a one-time setup of their repository for data backup, and connect an OAuth application to the instance of CLEF to enable user authentication (if needed). In turn, GitHub and Zenodo can be synchronised, associating a versioned DOI to each release. Likewise, web sources relevant to collected data can be flagged and automatically sent to the Internet Archive Wayback machine. These solutions allow us to close the circle of data FAIRness, ensuring long term preservation of digital archives. Currently the setup of GitHub and Zenodo is delegated to the user, who may decide not to publish their data, or to store data in a private repository.

\emph{Reusability and Sustainability.} Solutions like LED are not available for reuse, limiting their benefits to specific projects or user groups. CLEF, on the other hand, is available as an open source project for reuse, and an active community of Polifonia project members contribute to the development. 
The development of fully-fledged data management systems requires a significant amount of resources, skills, and time, which is often not sustainable in the long-term maintenance of a project. Rather, CLEF economically relies on established and popular solutions, therefore minimising the effort in maintaining the software to its core functionalities and improving its sustainability over time.

CLEF has a few known limitations. First, CLEF has so far been used by the members of ARTchives and Polifonia, and the development of CLEF has been focused on user-friendliness for editors. However, user-friendliness for end users is also important, in particular given that CLEF is intended for use by lay persons. Currently, the default configuration of CLEF allows end users to search records via text search and by means of filtering mechanisms. Tests with undergraduates are planned in the next months and will show how well it performs when used by such persons, and what are additional User Interface requirements to improve user-friendliness. 
Second, CLEF benefits greatly from a tight integration with Github. Not all cultural institutions and scholarly projects are, however, familiar with Github. It remains to be seen whether this is a barrier to uptake.
Third, sharing CLEF as an open source project provides a valuable service to cultural institutions and projects. However, it also comes with a responsibility. We must ensure that CLEF remains in active development, to at the very least ensure that it continues to work and does not leave its users vulnerable to security issues. The project is maintained by a community of institutions involved in Polifonia, including research centres, public bodies (e.g. the Italian Ministry of Cultural Heritage), and scholars. The consortium ensures development for the next 2 years, and the Digital Humanities Advanced Research Centre (/DH.arc), University of Bologna, ensures its maintenance in the long run.

Other limitations are due to pitfalls of technological choices. A particular issue of interest is the reuse of ontologies, which are reused via links to external services, rather than by importing them. This avoids duplication and update issues, and allows CLEF to benefit from development work and expertise of third parties. But it also creates a dependency, potentially causing issues if those services become unavailable, either temporarily or permanently. Changes to externally hosted ontologies could also have negative effects on the data quality, if the meaning of terms is changed. This is a well-known problem in the ontology engineering community \cite{carriero2020landscape}, which has not yet agreed on a shareable policy. 
\color{red}
Lastly, there are several benefits in relying on LOD principles, e.g. it allows for decoupling with no hard dependencies and seamless integration between systems that can link to each other via data URIs. Nonetheless, mechanisms for migrating data from museums legacy content management systems to the triplestore of CLEF are not implemented at the moment. In future works we will address this issue to allow a smooth transition.

\color{black}

\section{Conclusion}\label{conclusion}
In the Introduction, we emphasized the importance to cultural heritage applications of user-friendliness, provenance management, and integration with existing data management workflows to ensure reusability and sustainability. Via the use cases, we distilled these broad principles into specific requirements, namely reusability, accessibility, continuity, findability and preservation. In the previous section, we demonstrated how CLEF answers each of these requirements. This enables CLEF to support the gathering of cultural heritage data in a user-friendly manner while ensuring good provenance. In addition, CLEF leverages the benefits of Linked Open Data to ensure FAIRness of the data, encouraging reuse. CLEF has been designed to be integrated with common data management workflows. In particular, the integration with GitHub allows adopters to maximise data reuse by offering, along with a SPARQL endpoint, an automatically generated, versioned, data dump, that can be easily linked to Zenodo and assigned a DOI, ensuring its long-term preservation and citability. Finally, CLEF is available as an open source project, making it reusable, thus enabling good quality data gathering for cultural institutions and scholarly projects not in possession of the skills or resources to develop such tools themselves.

Future work will address known limitations and design improvements, driven by user studies with a broader audience. These include tests on usability of interfaces and development of mechanisms for automatic GitHub integration with CLEF and with local storage systems. Moreover, the agile and easy to integrate code base of CLEF offers the opportunity to experiment along new research lines. In particular, methods to discover new web resources based on the ones already included in catalogues (e.g. online music resources in musoW), and methods to enrich existing records with new information (e.g. auto-fill in of fields describing contents of archival collections) will be developed to support adopters in their daily tasks. Lastly, to fill the gap in data preservation and close the data life-cycle, methods to export and donate data to Wikidata will be implemented. 

\section{Acknowledgments}

This work is supported by a project that has received funding from the European Union’s Horizon 2020 research and innovation programme under grant agreement No 101004746 (Polifonia: a digital harmoniser for musical heritage knowledge, H2020-SC6-TRANSFORMATIONS).

\textbf{Authors' responsibilities} The authors collaborated in the design and the research. M. Daquino is responsible for section \ref{lodnative}, \ref{related}, and \ref{ARTchives}; E. Daga and M. Wigham are responsible for section \ref{musoW}, \ref{discussion} and \ref{conclusion}; F. Tomasi for section \ref{intro}. 

\bibliographystyle{ACM-Reference-Format}
\bibliography{main}


\begin{thebibliography}{42}


\ifx \showCODEN    \undefined \def \showCODEN     #1{\unskip}     \fi
\ifx \showDOI      \undefined \def \showDOI       #1{#1}\fi
\ifx \showISBNx    \undefined \def \showISBNx     #1{\unskip}     \fi
\ifx \showISBNxiii \undefined \def \showISBNxiii  #1{\unskip}     \fi
\ifx \showISSN     \undefined \def \showISSN      #1{\unskip}     \fi
\ifx \showLCCN     \undefined \def \showLCCN      #1{\unskip}     \fi
\ifx \shownote     \undefined \def \shownote      #1{#1}          \fi
\ifx \showarticletitle \undefined \def \showarticletitle #1{#1}   \fi
\ifx \showURL      \undefined \def \showURL       {\relax}        \fi
\providecommand\bibfield[2]{#2}
\providecommand\bibinfo[2]{#2}
\providecommand\natexlab[1]{#1}
\providecommand\showeprint[2][]{arXiv:#2}

\bibitem[spa(2013)]%
        {sparql11}
 \bibinfo{year}{2013}\natexlab{}.
\newblock \bibinfo{booktitle}{\emph{{SPARQL} 1.1 Overview}}.
\newblock \bibinfo{type}{{W3C} Recommendation}. \bibinfo{institution}{W3C}.
\newblock
\newblock
\shownote{https://www.w3.org/TR/2013/REC-sparql11-overview-20130321/}.


\bibitem[Adamou et~al\mbox{.}(2019)]%
        {adamou2019crowdsourcing}
\bibfield{author}{\bibinfo{person}{Alessandro Adamou}, \bibinfo{person}{Simon
  Brown}, \bibinfo{person}{Helen Barlow}, \bibinfo{person}{Carlo Allocca},
  {and} \bibinfo{person}{Mathieu d’Aquin}.} \bibinfo{year}{2019}\natexlab{}.
\newblock \showarticletitle{Crowdsourcing Linked Data on listening experiences
  through reuse and enhancement of library data}.
\newblock \bibinfo{journal}{\emph{International Journal on Digital Libraries}}
  \bibinfo{volume}{20}, \bibinfo{number}{1} (\bibinfo{year}{2019}),
  \bibinfo{pages}{61--79}.
\newblock


\bibitem[Amin et~al\mbox{.}(2008)]%
        {amin2008understanding}
\bibfield{author}{\bibinfo{person}{Alia Amin}, \bibinfo{person}{Jacco
  Van~Ossenbruggen}, \bibinfo{person}{Lynda Hardman}, {and}
  \bibinfo{person}{Annelies van Nispen}.} \bibinfo{year}{2008}\natexlab{}.
\newblock \showarticletitle{Understanding cultural heritage experts'
  information seeking needs}. In \bibinfo{booktitle}{\emph{Proceedings of the
  8th ACM/IEEE-CS joint conference on Digital libraries}}.
  \bibinfo{pages}{39--47}.
\newblock


\bibitem[Bonacchi et~al\mbox{.}(2019)]%
        {bonacchi2019participation}
\bibfield{author}{\bibinfo{person}{Chiara Bonacchi}, \bibinfo{person}{Andrew
  Bevan}, \bibinfo{person}{Adi Keinan-Schoonbaert}, \bibinfo{person}{Daniel
  Pett}, {and} \bibinfo{person}{Jennifer Wexler}.}
  \bibinfo{year}{2019}\natexlab{}.
\newblock \showarticletitle{Participation in heritage crowdsourcing}.
\newblock \bibinfo{journal}{\emph{Museum Management and Curatorship}}
  \bibinfo{volume}{34}, \bibinfo{number}{2} (\bibinfo{year}{2019}),
  \bibinfo{pages}{166--182}.
\newblock


\bibitem[Carriero(2020)]%
        {carriero2020landscape}
\bibfield{author}{\bibinfo{person}{Valentina et~al. Carriero}.}
  \bibinfo{year}{2020}\natexlab{}.
\newblock \showarticletitle{The landscape of ontology reuse approaches}.
\newblock \bibinfo{journal}{\emph{Applications and Practices in Ontology
  Design, Extraction, and Reasoning}}  \bibinfo{volume}{49}
  (\bibinfo{year}{2020}), \bibinfo{pages}{21}.
\newblock


\bibitem[Carroll et~al\mbox{.}(2005)]%
        {carroll2005named}
\bibfield{author}{\bibinfo{person}{Jeremy~J Carroll},
  \bibinfo{person}{Christian Bizer}, \bibinfo{person}{Pat Hayes}, {and}
  \bibinfo{person}{Patrick Stickler}.} \bibinfo{year}{2005}\natexlab{}.
\newblock \showarticletitle{Named graphs}.
\newblock \bibinfo{journal}{\emph{Journal of Web Semantics}}
  \bibinfo{volume}{3}, \bibinfo{number}{4} (\bibinfo{year}{2005}),
  \bibinfo{pages}{247--267}.
\newblock


\bibitem[Ceolin et~al\mbox{.}(2016)]%
        {ceolin2016combining}
\bibfield{author}{\bibinfo{person}{Davide Ceolin}, \bibinfo{person}{Paul
  Groth}, \bibinfo{person}{Valentina Maccatrozzo}, \bibinfo{person}{Wan
  Fokkink}, \bibinfo{person}{Willem Robert~Van Hage}, {and}
  \bibinfo{person}{Archana Nottamkandath}.} \bibinfo{year}{2016}\natexlab{}.
\newblock \showarticletitle{Combining user reputation and provenance analysis
  for trust assessment}.
\newblock \bibinfo{journal}{\emph{Journal of Data and Information Quality
  (JDIQ)}} \bibinfo{volume}{7}, \bibinfo{number}{1-2} (\bibinfo{year}{2016}),
  \bibinfo{pages}{1--28}.
\newblock


\bibitem[Consortium et~al\mbox{.}(2014)]%
        {world2014best}
\bibfield{author}{\bibinfo{person}{World Wide~Web Consortium} {et~al\mbox{.}}}
  \bibinfo{year}{2014}\natexlab{}.
\newblock \showarticletitle{Best practices for publishing linked data}.
\newblock  (\bibinfo{year}{2014}).
\newblock


\bibitem[Daga et~al\mbox{.}(2021)]%
        {daga2021integrating}
\bibfield{author}{\bibinfo{person}{Enrico Daga}, \bibinfo{person}{Luigi
  Asprino}, \bibinfo{person}{Marilena Daquino}, \bibinfo{person}{Rossana
  Damiano}, \bibinfo{person}{Belen Diaz~Agudo}, \bibinfo{person}{Aldo Gangemi},
  \bibinfo{person}{Tsvi Kuflik}, \bibinfo{person}{Antonio Lieto},
  \bibinfo{person}{Anna~Maria Marras}, \bibinfo{person}{Delfina
  Martinez~Pandiani}, \bibinfo{person}{Paul Mulholland}, {et~al\mbox{.}}}
  \bibinfo{year}{2021}\natexlab{}.
\newblock \showarticletitle{Integrating citizen experiences in cultural
  heritage archives: requirements, state of the art, and challenges}.
\newblock \bibinfo{journal}{\emph{Journal on Computing and Cultural Heritage
  (JOCCH)}} (\bibinfo{year}{2021}), \bibinfo{pages}{In--Press}.
\newblock


\bibitem[Daquino(2021)]%
        {daquino2021linked}
\bibfield{author}{\bibinfo{person}{Marilena Daquino}.}
  \bibinfo{year}{2021}\natexlab{}.
\newblock \showarticletitle{Linked Open Data native cataloguing and archival
  description}.
\newblock \bibinfo{journal}{\emph{Linked Open Data native cataloguing and
  archival description}} (\bibinfo{year}{2021}), \bibinfo{pages}{91--104}.
\newblock


\bibitem[Daquino et~al\mbox{.}(2017a)]%
        {daquino2017characterizing}
\bibfield{author}{\bibinfo{person}{Marilena Daquino}, \bibinfo{person}{Enrico
  Daga}, \bibinfo{person}{Mathieu d'Aquin}, \bibinfo{person}{Aldo Gangemi},
  \bibinfo{person}{Simon Holland}, \bibinfo{person}{Robin Laney},
  \bibinfo{person}{Albert~Merono Penuela}, {and} \bibinfo{person}{Paul
  Mulholland}.} \bibinfo{year}{2017}\natexlab{a}.
\newblock \showarticletitle{Characterizing the landscape of musical data on the
  Web: State of the art and challenges}.
\newblock  (\bibinfo{year}{2017}).
\newblock


\bibitem[Daquino et~al\mbox{.}(2021)]%
        {daquino2021artchives}
\bibfield{author}{\bibinfo{person}{Marilena Daquino}, \bibinfo{person}{Lucia
  Giagnolini}, {and} \bibinfo{person}{Francesca Tomasi}.}
  \bibinfo{year}{2021}\natexlab{}.
\newblock \showarticletitle{ARTchives: a Linked Open Data Native Catalogue of
  Art Historians’ Archives}. In \bibinfo{booktitle}{\emph{Theory and Practice
  of Digital Libraries}}.
\newblock


\bibitem[Daquino and Hlosta(2022)]%
        {marilena_daquino_2021_5603223}
\bibfield{author}{\bibinfo{person}{Marilena Daquino} {and}
  \bibinfo{person}{Martin Hlosta}.} \bibinfo{year}{2022}\natexlab{}.
\newblock \bibinfo{booktitle}{\emph{polifonia-project/clef: Revised CLEF}}.
\newblock
\urldef\tempurl%
\url{https://doi.org/10.5281/zenodo.6423933}
\showDOI{\tempurl}


\bibitem[Daquino et~al\mbox{.}(2017b)]%
        {daquino2017enhancing}
\bibfield{author}{\bibinfo{person}{Marilena Daquino},
  \bibinfo{person}{Francesca Mambelli}, \bibinfo{person}{Silvio Peroni},
  \bibinfo{person}{Francesca Tomasi}, {and} \bibinfo{person}{Fabio Vitali}.}
  \bibinfo{year}{2017}\natexlab{b}.
\newblock \showarticletitle{Enhancing semantic expressivity in the cultural
  heritage domain: exposing the Zeri Photo Archive as Linked Open Data}.
\newblock \bibinfo{journal}{\emph{Journal on Computing and Cultural Heritage
  (JOCCH)}} \bibinfo{volume}{10}, \bibinfo{number}{4} (\bibinfo{year}{2017}),
  \bibinfo{pages}{1--21}.
\newblock


\bibitem[Davis and Heravi(2021)]%
        {davis2021linked}
\bibfield{author}{\bibinfo{person}{Edie Davis} {and} \bibinfo{person}{Bahareh
  Heravi}.} \bibinfo{year}{2021}\natexlab{}.
\newblock \showarticletitle{Linked Data and Cultural Heritage: A Systematic
  Review of Participation, Collaboration, and Motivation}.
\newblock \bibinfo{journal}{\emph{Journal on Computing and Cultural Heritage
  (JOCCH)}} \bibinfo{volume}{14}, \bibinfo{number}{2} (\bibinfo{year}{2021}),
  \bibinfo{pages}{1--18}.
\newblock


\bibitem[Davis(2019)]%
        {davis2019old}
\bibfield{author}{\bibinfo{person}{Kelly Davis}.}
  \bibinfo{year}{2019}\natexlab{}.
\newblock \showarticletitle{Old metadata in a new world: Standardizing the
  Getty Provenance Index for linked data}.
\newblock \bibinfo{journal}{\emph{Art Libraries Journal}} \bibinfo{volume}{44},
  \bibinfo{number}{4} (\bibinfo{year}{2019}), \bibinfo{pages}{162--166}.
\newblock


\bibitem[De~Nies et~al\mbox{.}(2013)]%
        {de2013git2prov}
\bibfield{author}{\bibinfo{person}{Tom De~Nies}, \bibinfo{person}{Sara
  Magliacane}, \bibinfo{person}{Ruben Verborgh}, \bibinfo{person}{Sam Coppens},
  \bibinfo{person}{Paul Groth}, \bibinfo{person}{Erik Mannens}, {and}
  \bibinfo{person}{Rik Van~de Walle}.} \bibinfo{year}{2013}\natexlab{}.
\newblock \showarticletitle{Git2PROV: Exposing Version Control System Content
  as W3C PROV.}. In \bibinfo{booktitle}{\emph{International Semantic Web
  Conference (Posters \& Demos)}}. \bibinfo{pages}{125--128}.
\newblock


\bibitem[Deliot(2014)]%
        {deliot2014publishing}
\bibfield{author}{\bibinfo{person}{Corine Deliot}.}
  \bibinfo{year}{2014}\natexlab{}.
\newblock \showarticletitle{Publishing the British National Bibliography as
  linked open data}.
\newblock \bibinfo{journal}{\emph{Catalogue \& Index}}  \bibinfo{volume}{174}
  (\bibinfo{year}{2014}), \bibinfo{pages}{13--18}.
\newblock


\bibitem[Delmas-Glass and Sanderson(2020)]%
        {delmas2020fostering}
\bibfield{author}{\bibinfo{person}{Emmanuelle Delmas-Glass} {and}
  \bibinfo{person}{Robert Sanderson}.} \bibinfo{year}{2020}\natexlab{}.
\newblock \showarticletitle{Fostering a community of PHAROS scholars through
  the adoption of open standards}.
\newblock \bibinfo{journal}{\emph{Art Libraries Journal}} \bibinfo{volume}{45},
  \bibinfo{number}{1} (\bibinfo{year}{2020}), \bibinfo{pages}{19--23}.
\newblock


\bibitem[Dijkshoorn et~al\mbox{.}(2017)]%
        {dijkshoorn2017accurator}
\bibfield{author}{\bibinfo{person}{Chris Dijkshoorn}, \bibinfo{person}{Victor
  De~Boer}, \bibinfo{person}{Lora Aroyo}, {and} \bibinfo{person}{Guus
  Schreiber}.} \bibinfo{year}{2017}\natexlab{}.
\newblock \showarticletitle{Accurator: Nichesourcing for cultural heritage}.
\newblock \bibinfo{journal}{\emph{arXiv preprint arXiv:1709.09249}}
  (\bibinfo{year}{2017}).
\newblock


\bibitem[Dijkshoorn et~al\mbox{.}(2018)]%
        {dijkshoorn2018rijksmuseum}
\bibfield{author}{\bibinfo{person}{Chris Dijkshoorn}, \bibinfo{person}{Lizzy
  Jongma}, \bibinfo{person}{Lora Aroyo}, \bibinfo{person}{Jacco
  Van~Ossenbruggen}, \bibinfo{person}{Guus Schreiber}, \bibinfo{person}{Wesley
  Ter~Weele}, {and} \bibinfo{person}{Jan Wielemaker}.}
  \bibinfo{year}{2018}\natexlab{}.
\newblock \showarticletitle{The Rijksmuseum collection as linked data}.
\newblock \bibinfo{journal}{\emph{Semantic Web}} \bibinfo{volume}{9},
  \bibinfo{number}{2} (\bibinfo{year}{2018}), \bibinfo{pages}{221--230}.
\newblock


\bibitem[Diulio et~al\mbox{.}(2021)]%
        {diulio2021usability}
\bibfield{author}{\bibinfo{person}{Mar{\'\i}a de la~Paz Diulio},
  \bibinfo{person}{Juan~Cruz Gardey}, \bibinfo{person}{Anal{\'\i}a~Fernanda
  Gomez}, {and} \bibinfo{person}{Alejandra Garrido}.}
  \bibinfo{year}{2021}\natexlab{}.
\newblock \showarticletitle{Usability of data-oriented user interfaces for
  cultural heritage: A systematic mapping study}.
\newblock \bibinfo{journal}{\emph{Journal of Information Science}}
  (\bibinfo{year}{2021}), \bibinfo{pages}{01655515211001787}.
\newblock


\bibitem[Doerr et~al\mbox{.}(2010)]%
        {doerr2010europeana}
\bibfield{author}{\bibinfo{person}{Martin Doerr}, \bibinfo{person}{Stefan
  Gradmann}, \bibinfo{person}{Steffen Hennicke}, \bibinfo{person}{Antoine
  Isaac}, \bibinfo{person}{Carlo Meghini}, {and} \bibinfo{person}{Herbert
  Van~de Sompel}.} \bibinfo{year}{2010}\natexlab{}.
\newblock \showarticletitle{The europeana data model (edm)}. In
  \bibinfo{booktitle}{\emph{World Library and Information Congress: 76th IFLA
  general conference and assembly}}, Vol.~\bibinfo{volume}{10}.
  \bibinfo{pages}{15}.
\newblock


\bibitem[Haase et~al\mbox{.}(2019)]%
        {haase2019metaphactory}
\bibfield{author}{\bibinfo{person}{Peter Haase}, \bibinfo{person}{Daniel~M
  Herzig}, \bibinfo{person}{Artem Kozlov}, \bibinfo{person}{Andriy Nikolov},
  {and} \bibinfo{person}{Johannes Trame}.} \bibinfo{year}{2019}\natexlab{}.
\newblock \showarticletitle{metaphactory: A platform for knowledge graph
  management}.
\newblock \bibinfo{journal}{\emph{Semantic Web}} \bibinfo{volume}{10},
  \bibinfo{number}{6} (\bibinfo{year}{2019}), \bibinfo{pages}{1109--1125}.
\newblock


\bibitem[Hauswirth et~al\mbox{.}(2017)]%
        {hauswirth2017linked}
\bibfield{author}{\bibinfo{person}{Manfred Hauswirth}, \bibinfo{person}{Marcin
  Wylot}, \bibinfo{person}{Martin Grund}, \bibinfo{person}{Paul Groth}, {and}
  \bibinfo{person}{Philippe Cudr{\'e}-Mauroux}.}
  \bibinfo{year}{2017}\natexlab{}.
\newblock \showarticletitle{Linked data management}.
\newblock In \bibinfo{booktitle}{\emph{Handbook of big data technologies}}.
  \bibinfo{publisher}{Springer}, \bibinfo{pages}{307--338}.
\newblock


\bibitem[Heath and Bizer(2011)]%
        {heath2011linked}
\bibfield{author}{\bibinfo{person}{Tom Heath} {and} \bibinfo{person}{Christian
  Bizer}.} \bibinfo{year}{2011}\natexlab{}.
\newblock \showarticletitle{Linked data: Evolving the web into a global data
  space}.
\newblock \bibinfo{journal}{\emph{Synthesis lectures on the semantic web:
  theory and technology}} \bibinfo{volume}{1}, \bibinfo{number}{1}
  (\bibinfo{year}{2011}), \bibinfo{pages}{1--136}.
\newblock


\bibitem[Hermon and Niccolucci(2021)]%
        {hermon2021fair}
\bibfield{author}{\bibinfo{person}{Sorin Hermon} {and} \bibinfo{person}{Franco
  Niccolucci}.} \bibinfo{year}{2021}\natexlab{}.
\newblock \showarticletitle{FAIR Data and Cultural Heritage Special Issue
  Editorial Note}.
\newblock \bibinfo{journal}{\emph{International Journal on Digital Libraries}}
  \bibinfo{volume}{22}, \bibinfo{number}{3} (\bibinfo{year}{2021}),
  \bibinfo{pages}{251--255}.
\newblock


\bibitem[Herrera et~al\mbox{.}(2019)]%
        {herrera2019btc}
\bibfield{author}{\bibinfo{person}{Jos{\'e}-Miguel Herrera},
  \bibinfo{person}{Aidan Hogan}, {and} \bibinfo{person}{Tobias K{\"a}fer}.}
  \bibinfo{year}{2019}\natexlab{}.
\newblock \showarticletitle{BTC-2019: the 2019 billion triple challenge
  dataset}. In \bibinfo{booktitle}{\emph{International Semantic Web
  Conference}}. Springer, \bibinfo{pages}{163--180}.
\newblock


\bibitem[Knoblock et~al\mbox{.}(2017)]%
        {knoblock2017lessons}
\bibfield{author}{\bibinfo{person}{Craig~A Knoblock}, \bibinfo{person}{Pedro
  Szekely}, \bibinfo{person}{Eleanor Fink}, \bibinfo{person}{Duane Degler},
  \bibinfo{person}{David Newbury}, \bibinfo{person}{Robert Sanderson},
  \bibinfo{person}{Kate Blanch}, \bibinfo{person}{Sara Snyder},
  \bibinfo{person}{Nilay Chheda}, \bibinfo{person}{Nimesh Jain},
  {et~al\mbox{.}}} \bibinfo{year}{2017}\natexlab{}.
\newblock \showarticletitle{Lessons learned in building linked data for the
  American art collaborative}. In \bibinfo{booktitle}{\emph{International
  Semantic Web Conference}}. Springer, \bibinfo{pages}{263--279}.
\newblock


\bibitem[Koukopoulos et~al\mbox{.}(2017)]%
        {koukopoulos2017trustworthy}
\bibfield{author}{\bibinfo{person}{Zois Koukopoulos},
  \bibinfo{person}{Dimitrios Koukopoulos}, {and} \bibinfo{person}{Jason~J
  Jung}.} \bibinfo{year}{2017}\natexlab{}.
\newblock \showarticletitle{A trustworthy multimedia participatory platform for
  cultural heritage management in smart city environments}.
\newblock \bibinfo{journal}{\emph{Multimedia Tools and Applications}}
  \bibinfo{volume}{76}, \bibinfo{number}{24} (\bibinfo{year}{2017}),
  \bibinfo{pages}{25943--25981}.
\newblock


\bibitem[Lamprecht et~al\mbox{.}(2020)]%
        {lamprecht2020towards}
\bibfield{author}{\bibinfo{person}{Anna-Lena Lamprecht}, \bibinfo{person}{Leyla
  Garcia}, \bibinfo{person}{Mateusz Kuzak}, \bibinfo{person}{Carlos Martinez},
  \bibinfo{person}{Ricardo Arcila}, \bibinfo{person}{Eva Martin Del~Pico},
  \bibinfo{person}{Victoria Dominguez Del~Angel}, \bibinfo{person}{Stephanie
  Van De~Sandt}, \bibinfo{person}{Jon Ison}, \bibinfo{person}{Paula~Andrea
  Martinez}, {et~al\mbox{.}}} \bibinfo{year}{2020}\natexlab{}.
\newblock \showarticletitle{Towards FAIR principles for research software}.
\newblock \bibinfo{journal}{\emph{Data Science}} \bibinfo{volume}{3},
  \bibinfo{number}{1} (\bibinfo{year}{2020}), \bibinfo{pages}{37--59}.
\newblock


\bibitem[Lebo et~al\mbox{.}(2013)]%
        {lebo2013prov}
\bibfield{author}{\bibinfo{person}{Timothy Lebo}, \bibinfo{person}{Satya
  Sahoo}, \bibinfo{person}{Deborah McGuinness}, \bibinfo{person}{Khalid
  Belhajjame}, \bibinfo{person}{James Cheney}, \bibinfo{person}{David Corsar},
  \bibinfo{person}{Daniel Garijo}, \bibinfo{person}{Stian Soiland-Reyes},
  \bibinfo{person}{Stephan Zednik}, {and} \bibinfo{person}{Jun Zhao}.}
  \bibinfo{year}{2013}\natexlab{}.
\newblock \showarticletitle{Prov-o: The prov ontology}.
\newblock  (\bibinfo{year}{2013}).
\newblock


\bibitem[Malmsten(2009)]%
        {malmsten2009exposing}
\bibfield{author}{\bibinfo{person}{Martin Malmsten}.}
  \bibinfo{year}{2009}\natexlab{}.
\newblock \showarticletitle{Exposing library data as linked data}.
\newblock \bibinfo{journal}{\emph{IFLA satellite preconference sponsored by the
  Information Technology Section" Emerging trends in}} (\bibinfo{year}{2009}).
\newblock


\bibitem[Mihindukulasooriya et~al\mbox{.}(2013)]%
        {mihindukulasooriya2013linked}
\bibfield{author}{\bibinfo{person}{Nandana Mihindukulasooriya},
  \bibinfo{person}{Raul Garcia-Castro}, {and} \bibinfo{person}{Miguel~Esteban
  Guti{\'e}rrez}.} \bibinfo{year}{2013}\natexlab{}.
\newblock \showarticletitle{Linked Data Platform as a novel approach for
  Enterprise Application Integration.}. In \bibinfo{booktitle}{\emph{COLD}}.
\newblock


\bibitem[Moreau(2010)]%
        {moreau2010foundations}
\bibfield{author}{\bibinfo{person}{Luc Moreau}.}
  \bibinfo{year}{2010}\natexlab{}.
\newblock \bibinfo{booktitle}{\emph{The foundations for provenance on the
  web}}.
\newblock \bibinfo{publisher}{Now Publishers Inc}.
\newblock


\bibitem[Orlandi et~al\mbox{.}(2021)]%
        {orlandi2021benchmarking}
\bibfield{author}{\bibinfo{person}{Fabrizio Orlandi}, \bibinfo{person}{Damien
  Graux}, {and} \bibinfo{person}{Declan O'Sullivan}.}
  \bibinfo{year}{2021}\natexlab{}.
\newblock \showarticletitle{Benchmarking RDF Metadata Representations:
  Reification, Singleton Property and RDF}. In \bibinfo{booktitle}{\emph{2021
  IEEE 15th International Conference on Semantic Computing (ICSC)}}. IEEE,
  \bibinfo{pages}{233--240}.
\newblock


\bibitem[R{\"o}der et~al\mbox{.}(2020)]%
        {roder2020hobbit}
\bibfield{author}{\bibinfo{person}{Michael R{\"o}der}, \bibinfo{person}{Denis
  Kuchelev}, {and} \bibinfo{person}{Axel-Cyrille Ngonga~Ngomo}.}
  \bibinfo{year}{2020}\natexlab{}.
\newblock \showarticletitle{HOBBIT: A platform for benchmarking Big Linked
  Data}.
\newblock \bibinfo{journal}{\emph{Data Science}} \bibinfo{volume}{3},
  \bibinfo{number}{1} (\bibinfo{year}{2020}), \bibinfo{pages}{15--35}.
\newblock


\bibitem[Sandusky(2016)]%
        {sandusky2016computational}
\bibfield{author}{\bibinfo{person}{Robert~J Sandusky}.}
  \bibinfo{year}{2016}\natexlab{}.
\newblock \showarticletitle{Computational provenance: Dataone and implications
  for cultural heritage institutions}. In \bibinfo{booktitle}{\emph{2016 IEEE
  International Conference on Big Data (Big Data)}}. IEEE,
  \bibinfo{pages}{3266--3271}.
\newblock


\bibitem[Sikos and Philp(2020)]%
        {sikos2020provenance}
\bibfield{author}{\bibinfo{person}{Leslie~F Sikos} {and} \bibinfo{person}{Dean
  Philp}.} \bibinfo{year}{2020}\natexlab{}.
\newblock \showarticletitle{Provenance-aware knowledge representation: A survey
  of data models and contextualized knowledge graphs}.
\newblock \bibinfo{journal}{\emph{Data Science and Engineering}}
  \bibinfo{volume}{5}, \bibinfo{number}{3} (\bibinfo{year}{2020}),
  \bibinfo{pages}{293--316}.
\newblock


\bibitem[Thompson et~al\mbox{.}(2016)]%
        {thompson2016bigdata}
\bibfield{author}{\bibinfo{person}{Bryan Thompson}, \bibinfo{person}{Mike
  Personick}, {and} \bibinfo{person}{Martyn Cutcher}.}
  \bibinfo{year}{2016}\natexlab{}.
\newblock \showarticletitle{The Bigdata{\textregistered} RDF graph database}.
\newblock In \bibinfo{booktitle}{\emph{Linked Data Management}}.
  \bibinfo{publisher}{Chapman and Hall/CRC}, \bibinfo{pages}{221--266}.
\newblock


\bibitem[Vandenbussche et~al\mbox{.}(2017)]%
        {vandenbussche2017linked}
\bibfield{author}{\bibinfo{person}{Pierre-Yves Vandenbussche},
  \bibinfo{person}{Ghislain~A Atemezing}, \bibinfo{person}{Mar{\'\i}a
  Poveda-Villal{\'o}n}, {and} \bibinfo{person}{Bernard Vatant}.}
  \bibinfo{year}{2017}\natexlab{}.
\newblock \showarticletitle{Linked Open Vocabularies (LOV): a gateway to
  reusable semantic vocabularies on the Web}.
\newblock \bibinfo{journal}{\emph{Semantic Web}} \bibinfo{volume}{8},
  \bibinfo{number}{3} (\bibinfo{year}{2017}), \bibinfo{pages}{437--452}.
\newblock


\bibitem[Wilkinson et~al\mbox{.}(2016)]%
        {wilkinson2016fair}
\bibfield{author}{\bibinfo{person}{Mark~D Wilkinson}, \bibinfo{person}{Michel
  Dumontier}, \bibinfo{person}{IJsbrand~Jan Aalbersberg},
  \bibinfo{person}{Gabrielle Appleton}, \bibinfo{person}{Myles Axton},
  \bibinfo{person}{Arie Baak}, \bibinfo{person}{Niklas Blomberg},
  \bibinfo{person}{Jan-Willem Boiten}, \bibinfo{person}{Luiz~Bonino da
  Silva~Santos}, \bibinfo{person}{Philip~E Bourne}, {et~al\mbox{.}}}
  \bibinfo{year}{2016}\natexlab{}.
\newblock \showarticletitle{The FAIR Guiding Principles for scientific data
  management and stewardship}.
\newblock \bibinfo{journal}{\emph{Scientific data}} \bibinfo{volume}{3},
  \bibinfo{number}{1} (\bibinfo{year}{2016}), \bibinfo{pages}{1--9}.
\newblock


\end{thebibliography}

\end{document}